\documentclass{article}
\usepackage{emulateapj}
\usepackage{psfig}

\slugcomment{}

\lefthead{Maiolino et al.}
\righthead{Nuclear gas bar in Circinus}

\begin{document}

\title{Discovery of a nuclear gas bar feeding the active nucleus in
Circinus\altaffilmark{1}}

\altaffiltext{1}{Based on observations with the NASA/ESA Hubble
Space Telescope (obtained at the Space Telescope Science Institute, which
is operated by the Association of Universities for Research in Astronomy,
Inc. under NASA contract No. NAS5-26555) and with the Anglo--Australian
Telescope (Sinding Spring, Australia).}

\author{R.~Maiolino\altaffilmark{2}, A.~Alonso-Herrero\altaffilmark{3},
 S.~Anders\altaffilmark{4}, A.~Quillen\altaffilmark{3},
 M.J.~Rieke\altaffilmark{3}, G.H.~Rieke\altaffilmark{3},
 L.E.~Tacconi-Garman\altaffilmark{4}}

\altaffiltext{2}{Osservatorio Astrofisico di Arcetri, L.go E. Fermi 5,
 I-50125, Firenze, Italy}
\altaffiltext{3}{Steward Observatory, The University of Arizona, 933
N. Cherry Ave., Tucson, AZ 85721}
\altaffiltext{4}{Max-Planck-Institut f\"{u}r Extraterrestrische Physik,
Postfach 1603, D-85740, Garching, Germany}

\begin{abstract}

We report the discovery of gas inflow motions towards the active
nucleus of the Circinus galaxy caused by the non-axisymmetric
potential of a nuclear gas bar. Evidence for dust associated with the
bar comes from the HST/NICMOS $H-K$ color map, whereas the streaming
motions along the gas bar are seen in the velocity field of
the H$_2$S(1)(1--0) emission line. The gas bar is about 100\,pc
long with a visual extinction in excess of 10\,mag. Indication for the
gaseous nature of this bar comes from the lack of a stellar counterpart
even in the $K$ band where the extinction is greatly reduced.

We also use the NICMOS emission line images (Pa$\alpha$,
[Si\,{\sc vi}], and [Fe\,{\sc ii}]) to
study the innermost region of the ionization cones
and the nuclear star forming activity. We discuss
the possible relationship of these components with the gaseous bar.

\end{abstract}


\section{Introduction} \label{intro}

Active galactic nuclei (AGNs) are thought to be powered 
by an accreting supermassive
black hole ($\rm M_{BH} \approx 10^6 -10^{10} M_{\odot}$).
The required accretion rate ranges from
$\rm 10^{-2}-10^{-1} ~M_{\odot} yr^{-1}$ (Seyfert nuclei) up to
$\rm 10-100 ~M_{\odot} yr^{-1}$ (QSOs). Although the gas content of host
galaxies is generally enough to fuel AGNs over the Hubble time,
the mechanism responsible for transporting gas from the kpc scale
into the nuclear region is still unclear. Indeed, the galactic gas must
lose about 99.9\% of its angular momentum before reaching the central
few parsecs, where viscosity can be effective. The required accretion rates
are lower in Seyfert galaxies than in QSOs and the gas inflow requirements
from the host galaxy are relaxed. However, because of the large number
of Seyfert galaxies ($\sim$ 10\%) in the local Universe and the even
larger fraction of AGNs at high redshifts
(Ho et al. 1997, Maiolino \& Rieke 1995, Hasinger et al. 1998, Miyaji et
al. 1999), some mechanism for transporting gas from the circumnuclear 100 pc
into the central parsec is still required to maintain the activity of
these nuclei.
Also, the accumulation of large quantities of gas
in the innermost region of Seyfert nuclei is inferred from the observation
that most of them are obscured by gaseous column densities in excess
of $\rm 10^{24} cm^{-2}$ within the central 10 pc (Risaliti et al. 1999)
and from the detection of maser emission within the nuclear few parsecs
(Greenhill et al. 1997a, 1997b, 1998b, Greenhill \& Gwinn 1997, Miyoshi
et al. 1995).

Non--axisymmetric gravitational potentials are
thought to be an efficient mechanism for transporting gas from the host galaxy
into the central region (Athanassoula 1992, Helfer \& Blitz 1995,
Laine et al. 1999, Hernquist 1989, Quillen et al. 1995, Regan et al. 1997,
Shlosman et al. 1990, Maiolino et al. 1999a). However,
within the central few hundred parsecs large scale disturbances have little
effect in removing angular momentum from the gas, since on these scales
the gravitational field is dominated by the central region
of the bulge that is axially symmetric. Nested, secondary stellar bars have
been observed in several barred system
(Jungwiert et al. 1997, Friedli et al. 1996,
Wozniak et al. 1995). These secondary bars are relatively small (a
few 100 pc long) and could be responsible for transporting gas into the
nuclear region to feed a black hole. Indeed, the fraction of Seyfert nuclei
among the 40 double-barred systems known so far is about 1/3 to be compared
to the fraction of Seyferts among all galaxies that is about 1/10
(Friedli 1999). Alternatively, Shlosman et al. (1989), Wada \& Habe (1992) and
Heller \& Shlosman (1994) proposed that if the mass of gas gathered into the
central region exceeds about 10\% of the dynamical mass then the gaseous disk
develops gravitational instabilities and forms a gaseous bar, which can
drive gas further into the nuclear region. Indeed, there is
evidence in several Seyfert nuclei for concentrations of gas whose mass
approaches the dynamical mass (Meixner et al. 1990, Henkel et al. 1991,
Risaliti et al. 1999), therefore gaseous bars are expected
to form and to provide a mechanisms for the fuelling of the nuclear region.
However, neither of these mechanisms appear to operate in the majority
of Seyfert galaxies.

Understanding the mechanism responsible for the fuelling of active galaxies
is also relevant to other issues. The dense gas funneled into the
central region is photoionized by UV radiation from the nuclear source 
and/or is shock excited by the interaction with radio jets.
Also, the large mass of gas driven into the central region and the
perturbations due to the fuelling mechanism are expected to boost the star
forming activity in the central region. Therefore, the fuelling mechanism
of AGNs is likely to play a key role in the connection between starbursts
and AGNs (Gonzalez-Delgado et al. 1998, Maiolino et al. 1997).

The Circinus galaxy is a nearby (4 Mpc), edge--on (incl.$\sim 65^\circ$),
Sb--d system that is seen through a low interstellar extinction
window near the Galactic plane (A$_V$ = 1.5 mag, Freeman et al. 1977).
The nuclear optical line ratios (Oliva et al. 1994) are typical of a
Seyfert 2 galaxy. This classification is supported by the detection of
intense coronal lines (Oliva et al. 1994, Moorwood et al. 1996), the
discovery of an intense X--ray iron 6.4 keV line (Matt et al. 1996),
rapid variation of
the powerful H$_2$O maser emission (Greenhill et al. 1997c) and a prominent
ionization cone in the [OIII]$\lambda$5007 maps (Marconi et al. 1994a) with
filamentary supersonic outflows (Veilleux \& Bland-Hawthorn 1997).
The detection of an X-ray excess above 30 keV reveals that the
central engine is heavily obscured by a gaseous column density
$\rm N_H =4 \times 10^{24} cm^{-2}$ along our line of sight (Matt et al. 1999).
Large amounts of molecular gas in the central region of this galaxy have been
inferred also from CO observations (Aalto et al. 1995, Elmouttie et al. 1998a,
Curran et al. 1998, Johansson et al. 1991) and from the detection of a
nuclear maser disk (Greenhill et al. 1998). H$\alpha$ and
[SII] narrow band images (Marconi et al. 1994a) have  
revealed a star forming ring at a radius
of $\sim$10$''$=200pc, while Maiolino et al. (1998) have
found evidence for a young stellar population in the central 10--100 pc
with typical ages of the order of 10$^8$ yr. As a consequence,
Circinus is an excellent candidate to search for a gaseous bar
that might be feeding both the black hole and the star forming activity
in the central region.  

In this paper we present high angular resolution near infrared HST images
and integral field spectra of the central region of the Circinus galaxy that
reveal the presence of a small nuclear gaseous bar and show evidence
for streaming motions of the gas along the bar.

\section{Observations and data reduction} \label{obs}

\subsection{NICMOS observations} \label{nic_obs}

{\it HST}/NICMOS observations of Circinus were
obtained during two different orbits on March 16 and October 16 1998
using the NIC2 and NIC3 cameras.
In Table~1 we give the log of the NICMOS observations.
Column~(1) lists the NICMOS camera; column~(2), the filter; column~(3),
the corresponding ground-based broad-band filter, emission line or adjacent
continuum; column~(4), the total integration time in seconds; and  column~(5),
the original orientation of the images. The observations were taken in
a spiral dither with a 5.5 pixel spacing between each of four  positions.
The plate scales for cameras NIC2 and NIC3 are 0.076"/pixel and 0.204"/pixel
respectively.

The reduction of the NICMOS images used routines from the package
NicRed (McLeod 1997). A master dark image was produced by
combining between 10 and 20 darks for a given sample sequence after the
subtraction of the first readout. The darks were taken from other programs
carried out close in time. When possible we also generated our own flatfield
images from on-orbit data. The flatfields used for the NIC3 narrow-band
images were obtained as part of the camera 3 campaign in January 1998
and were kindly reduced and provided by Dr. Rodger Thompson. Only for the
NIC2 F187N and NIC2 F190N filters were in-flight flats not available,
so thermal-vacuum flatfields were used instead. The data reduction was
performed using the following steps: subtraction of the first readout; dark
current subtraction on a read-by-read basis; correction for linearity and cosmic
ray rejection (using the {\it fullfit} routine in NicRed);
and flatfielding. The individual dithered galaxy images
were registered to a common position using fractional pixel offsets and a
cubic spline interpolation, and combined to produce the final image of each
field. Since our NICMOS images were obtained after August 1997, no correction
for the pedestal effect was necessary.

Prior to flux calibrating  the images, the background needs to be
subtracted from the images (this is important only for filters at wavelengths
longer than 2$\mu$m). Due to the large projected size of Circinus, the
field of view of the NIC2 images is not large enough to allow measurements
of the background on blank corners of the images, so we used background
measurements taken during the Servicing Mission Observatory Verification
(SMOV) program for the filter NIC2 F222M, and from our observations of
the interacting
galaxy Arp~299 for the NIC2 F160W filter. The NIC3 mosaics cover a larger
field of view, and an estimate of the thermal background was obtained
from the corners of the images. The flux calibration of the broad-band,
on-line, and off-line images was performed using conversion factors
(from ADU/s to mJy) from measurements of the standard star P330-E during
SMOV.

NICMOS only provides continuum bands to the red of the emission lines.
 The extinction in the nuclear regions of Circinus is
very high (even at infrared wavelengths), and a straight subtraction of
the longer wavelength off-line image results in an overcorrection of the
continuum at the wavelength of the emission line. To estimate the continuum at
$1.644\,\mu$m, $1.96\,\mu$m, $2.122\,\mu$m and $1.87\,\mu$m
for the NIC3 F164N, NIC3 F196N and NIC3 212N
respectively, we fit the continuum between
$1.66\,\mu$m and $1.90\,\mu$m using the NIC3 F166N and NIC3 F200N
line-free images for the first two, and the NIC3 F200N and NIC3 F215N
for the last image. The same procedure was followed to subtract the
continuum at $1.87\,\mu$m for the NIC2 F187N image, using the
NIC2 F160W and F222M images. The continuum fit is simply a linear
regression of the flux as a function of
the wavelength for each point in the image. An image at the
required continuum wavelength is then constructed by extrapolation
or interpolation of the corresponding fit. The extrapolated continuum
is subtracted from the line+continuum images to produce the final
[Fe\,{\sc ii}]$1.644\,\mu$m, Pa$\alpha$, [Si\,{\sc vi}]$1.96\,\mu$m and
H$_2$ at $2.121\,\mu$m  images. Due to the low surface brightness of the
H$_2$ emission together with problems associated with
the continuum subtraction, the resulting H$_2$ image had
relatively low signal to noise, and therefore we used the H$_2$ map
obtained with the 3D integral field spectrometer (see next section). Note
however that when we rebinned the NICMOS H$_2$ image to the angular resolution
of the 3D image, both images showed a similar morphology.

The images presented in this paper were taken during different orbits and hence have different orientations (see Table~1).
Image rotation was removed by linear interpolation to the conventional
orientation (north up, east to the left), both to provide the same
orientation for all the images, and also to allow easy comparisons with
previously published data. 

\begin{deluxetable}{lcccc}
\tablecaption{Log of the NICMOS observations.}
\tablehead{\colhead{Camera} & \colhead{Filter} & \colhead{line/broad/cont} &
\colhead{$t_{\rm exp}$} & \colhead{orient}}
\startdata
NIC2   & F160W & $H$ & 192 & 94\arcdeg \\
NIC2   & F222M & $K$ & 192 & 94\arcdeg \\
NIC2   & F187N & Pa$\alpha$ & 384 & $-64\arcdeg$ \\
NIC2   & F190N & Pa$\alpha$ cont. & 384 & $-64\arcdeg$ \\
NIC3   & F164N & [Fe\,{\sc ii}]$1.664\,\mu$m & 256 & 94\arcdeg\\
NIC3   & F166N & [Fe\,{\sc ii}] cont. & 256 & 94\arcdeg\\
NIC3   & F196N & [Si\,{\sc vi}]$1.96\,\mu$m & 320 & 94\arcdeg \\
NIC3   & F200N & [Si\,{\sc vi}] cont.  & 320 & 94\arcdeg \\
NIC3   & F212N & H$_2$ $2.122\,\mu$m & 640 & $-64\arcdeg$ \\
NIC3   & F215N & H$_2$ cont. & 640 & $-64\arcdeg$ \\
\enddata
\end{deluxetable}

\subsection{3D integral field spectroscopy} \label{3d_obs}

We observed the nuclear region of the Circinus galaxy on March 18$^{\rm th}$
and 19$^{\rm th}$ 1998 with the near infrared integral field spectrometer
3D (Weitzel et al. 1996), assisted by ROGUE, a first order adaptive optics
system (Thatte et al. 1995), at the Anglo Australian Telescope.
3D slices the focal plane into 16 slits and disperses their light in
wavelength; the spectrum of the slices is then imaged onto a NICMOS3
detector (256$\times$256 pixels). The resulting ``cube'' provides
simultaneous spectra of an area $6.4''\times 6.4''$, as projected on the
sky, divided into 16$\times$16 pixels (i.e. $0.4''$/pix). Circinus was
observed with a grism providing a spectral resolution of $\lambda/\Delta
\lambda = 2100$
and covering the spectral range from 1.95$\rm \mu m$ to 2.21$\rm \mu m$.
The average seeing for accepted observations, after tip-tilt correction,
was about $0''.7-0''.8$. The (limited) field of view was not centered on
the nucleus
but was moved about 1 arcsec to the south-east so that it could include
a dusty feature we were interested in. The total on source integration time
was 20 minutes, split in twelve elementary integrations of 100 seconds.
The elementary on-source integrations were interleaved with an equal number
of acquisitions of sky frames for background subtraction. To avoid and control
instrumental artifact we dithered the field of view around the region of
interest so that each region of the source
 was observed by different regions of the
slicer and of the detector.
We observed a nearby O9 star to correct the atmospheric
transmission features. Data reduction was performed as described in
Weitzel et al. (1996) and Thatte et al. (1997). The subtraction of the
continuum underneath the emission lines was performed with a first order
fit to the continuum approximately $\pm$ 0.03 $\mu$m from each line.

\section{The nuclear gas bar} \label{bar}

\subsection{Evidence for a radial dust lane} \label{nic_bar}

Fig.~1a
shows a false color image of the central region of the Circinus
galaxy ($26''\times 29'' = 530\times 590$ pc) obtained by combining
HST images at 0.606 $\rm \mu m$ (WFPC2-F606W, blue),
at 1.66 $\rm \mu m$ (NIC3-F166N, green), and at 2.0 $\rm \mu m$
(NIC3-F200N, red). The galactic disk is inclined by about 65$^{\circ}$
with respect to our line of sight (the South-East is the near side of
the disk) and the position angle of the major axis is about 25$^{\circ}$.
The image shows a spiral pattern with prominent dust lanes, consistent
with the large amount of gas in this region as inferred from the CO
millimetric data. Fig.~1b shows the H--K color map constructed
with the NIC2-F160W ($\sim$ H band)
and NIC2-F222M ($\sim$ K band) images. 

\begin{figure*}
\centerline{
\psfig{figure=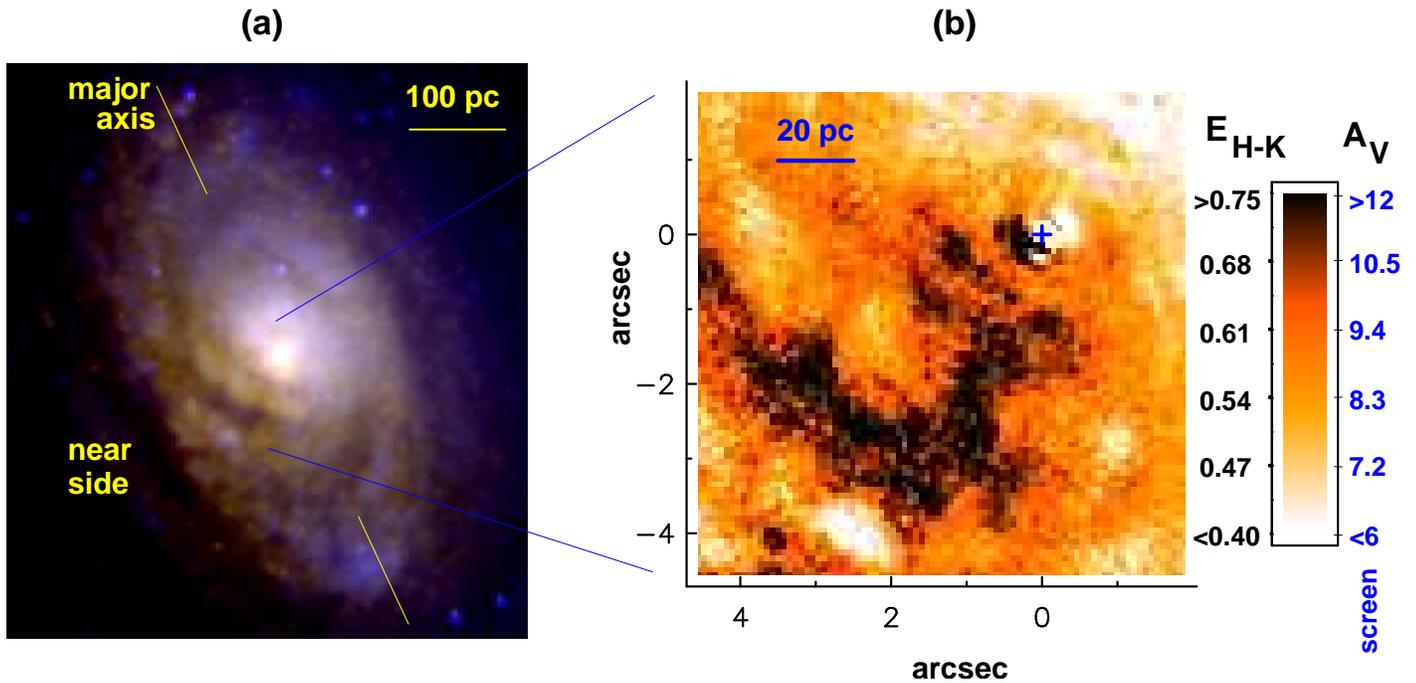,angle=-90,width=1.00\linewidth,clip=yes}}
\caption{\label{fig_hk} 
{\it a)} True color image of the central region of the Circinus
galaxy obtained by combining HST images at 0.606 $\mu$m (WFPC2-F606W, blue),
at 1.66 $\mu$m (NIC3-F166N, green) and at 2.0 $\mu$m (NIC3-F200N, red).
{\it b)} H--K color map of the nuclear region
constructed with the NIC2-F160W ($\sim$ H band)
and NIC2-F222M ($\sim$ K band) images. The nuclear PSF was subtracted in both
filters before producing the color map. The right hand side bar converts
the color code into H--K color excess and into visual extinction in the
case of a foreground screen model.
}
\end{figure*}

\begin{figure*}
\centerline{
\psfig{figure=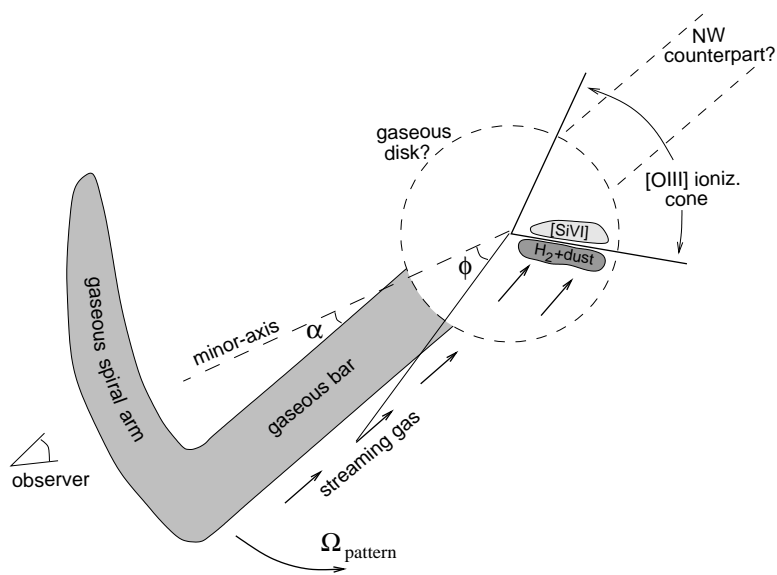,angle=0,width=0.55\linewidth,clip=yes}}
\caption{\label{fig_sketch} 
Schematic drawing of the geometry of the various components
within the central few 100 pc of Circinus.
}
\end{figure*}

The H--K color of the nucleus (identified with the 2$\rm \mu m$ peak) is
extremely red and in the F222M ($\sim$ K band)
image the nucleus dominates the emission within the central two arcseconds.
As discussed in detail in Maiolino et al. (1998), the powerful K band emission
is due to hot dust, heated by the active nucleus and located in a region
smaller than 1.5 pc (= 0$''$.07 arcsec) in radius. In Fig.~1b
we have subtracted a nuclear unresolved source using the NIC2 point spread
function as modeled by the {\it TinyTim} software (Krist \& Hook 1997).
This subtraction allows study of the H--K color of the circumnuclear two
arcseconds (although in the innermost arcsecond there are some residuals
probably due to imperfect subtraction). The PSF subtraction does not
affect the color map outside this inner region. 

\begin{figure*}
\centerline{
\psfig{figure=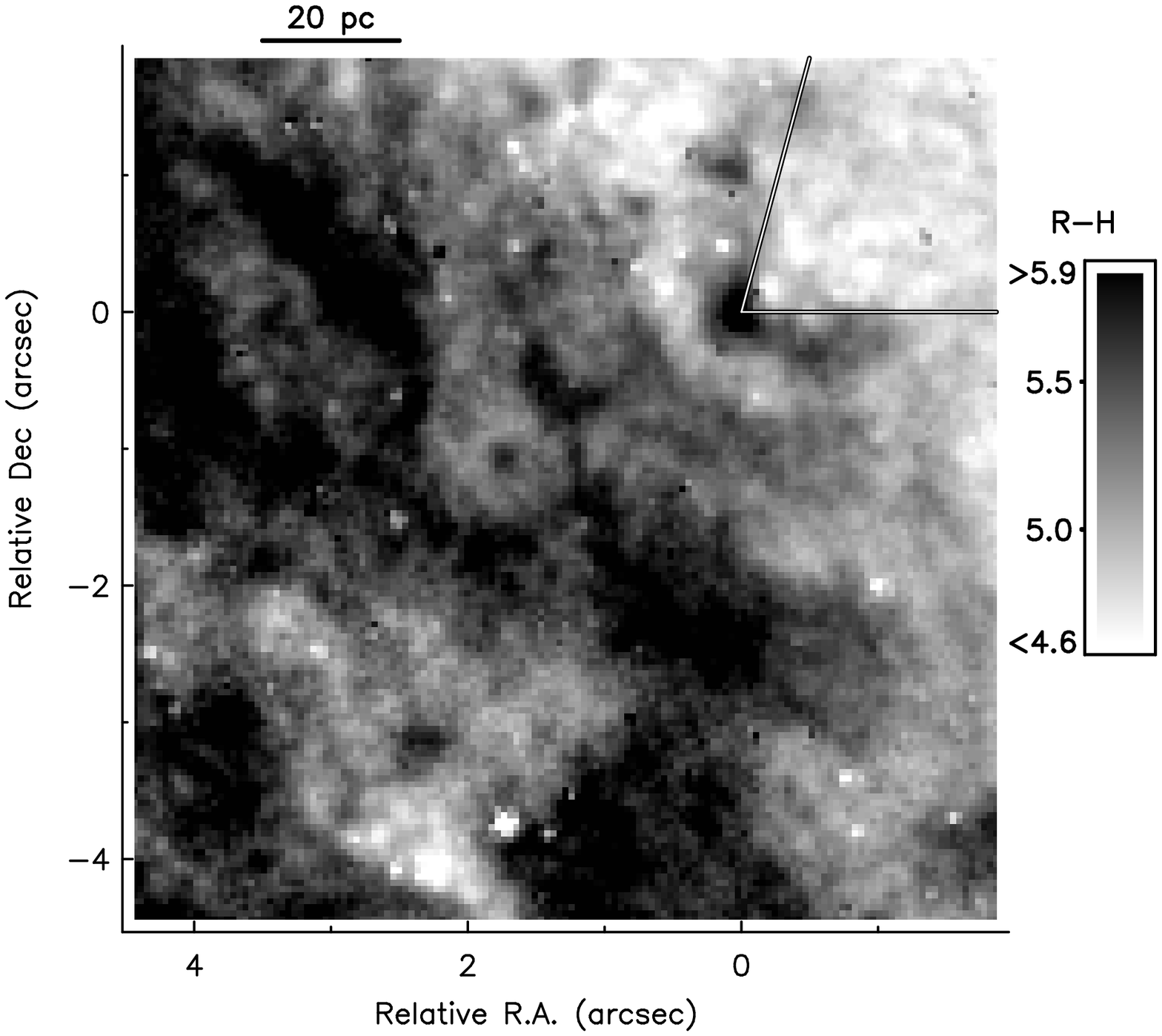,angle=0,width=0.55\linewidth,clip=yes}}
\caption{\label{fig_rh} 
R--H color map obtained by means of the WFPC2-F606W image
and the NIC2-F160W image. The straight
lines indicate the orientation of the light cone observed in [OIII].
}
\end{figure*}

The H--K color of the circumnuclear region is still significantly redder
than colors typical of stellar populations older than 10$^7$ yr:
(H--K)$\approx 0.22$, as observed both in galaxies that do not suffer
heavy extinction, and also as predicted by population models. Such off-nuclear
red colors cannot be ascribed to hot dust emission in the K band, for the
CO stellar bands at 2.29 $\mu$m do not appear diluted by non-stellar
emission in these regions (Maiolino et al. 1998). In addition, the H--K
color does not correspond with the Pa$\alpha$ and Br$\gamma$ emission
(see Sect.\ref{sf}), as expected if the UV continuum were heating dust
locally.  The only plausible explanation
is that the H--K color excess is due to reddening.

There is a moderate large-scale extinction gradient
across the minor axis of the galactic disk (i.e. from the NW to the SE),
resulting from the greater effectiveness of dust absorption on the
near side of the galactic disk (fewer bulge stars in front of us and more
bulge stars behind the dusty disk). However, the most prominent feature
of the ``extinction map'' in Fig.~1b
is the L-shaped feature extending from the
nucleus to the south-east\footnote{Evidence for this strucure, though at lower
angular resolution, is observed also in the H--K color map presented in
Maiolino et al. (1998).}: we believe that the section of this
feature elongated radially to the south is tracing
the near side of a nuclear gaseous-dusty bar.
The linear projected (half) size is about 50 pc.
The southern end of the dusty bar tilts by about 90 degrees to the east, in
a tightly wound spiral arm-like feature (see also Fig.~1a); this
morphology is similar to that observed in stellar bars on larger scales,
but in a gaseous version. A stellar counterpart is not present even in
the K band image, where the extinction is reduced, suggesting the bar
is purely gaseous.

An alternative explanation of the radial dust lane could be a warped disk
(eg. Quillen et al. 1993). However, as discussed in the next section,
the velocity field of the molecular gas strongly deviates from circular
motion, ruling out this interpretation.

The northern extremity of the gaseous bar ends in a dusty
feature that resembles a circumnuclear disk, about 20 pc in radius.
Only the near side of this disk-like feature is clearly delineated, possibly
as a consequence its being seen in absorption (as for the bar) and
therefore the contrast is expected to be much higher on the near side.
According to models, the gaseous bar is expected to form a gaseous disk
in its central part during the latest stages of its evolution as a
consequence of cloud-cloud viscous collisions at the denser regions near
the aphelia of the orbits (Wada \& Habe 1992).

Fig.~2
schematically shows the geometry of the gaseous features discussed
in this section along with other components discussed in the following
sections. The observed angle between the line of the minor axis and the
gaseous bar is $\rm \alpha _{\rm obs} \approx 30^{\circ}$, while its
deprojected value is $\rm \alpha = \tan ^{-1} (\tan \alpha _{\rm obs}
\cos {\it i}) = 16^{\circ}$, where {\it i} is the inclination angle of the
galaxy (65$^{\circ}$). The deprojected length of the bar is

\begin{equation} 
\rm L (bar) = L_{\rm obs}(bar) {\cos \alpha _{\rm obs} \over \cos {\it i}
\cdot \cos \alpha} \approx 105~pc
\end{equation}

We can use the H--K color excess to derive the extinction affecting the
nuclear region and, in particular, the extinction introduced by the dust
in the gas bar. By assuming that the intrinsic color of the stellar population
is (H--K)$_0=$0.22, we derive the magnitude of the dust absorption
affecting the stellar light. By adopting the extinction curve of
He et al. (1995) and a foreground dusty screen model we obtain:
$\rm E_{\rm H-K} = 0.065 ~ A_V $.
The color bar on the right-hand side of Fig.~1b
converts the color coding of the H--K map into color excess E$_{\rm H-K}$
and into visual extinction according to the extreme case of a foreground
dusty screen.

\begin{figure*}
\centerline{
\psfig{figure=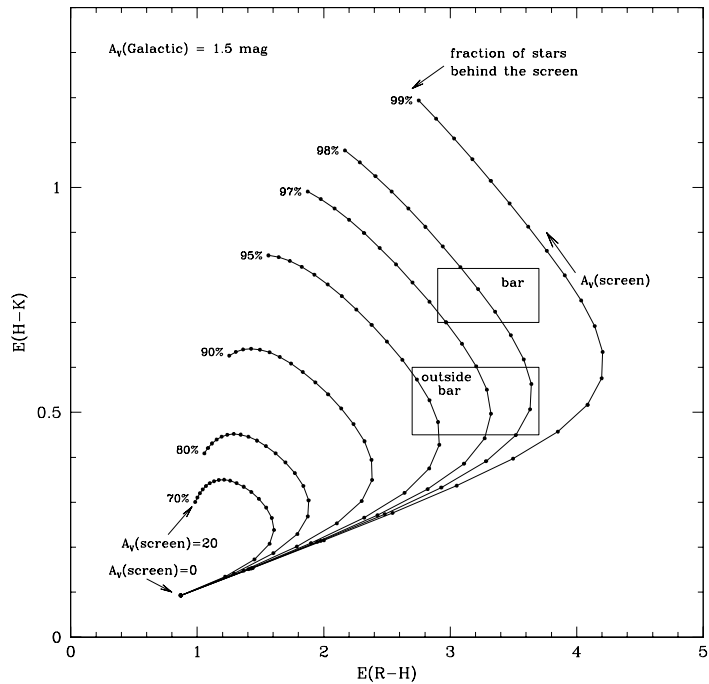,angle=0,width=0.5\linewidth,clip=yes}}
\caption{\label{fig_dustmod} 
Expected H--K and R--H color excess in the case of a dusty
screen that obscures only a fraction of the stars (the whole system is then
obscured by a Galactic extinction $\rm A_V = 1.5$ mag). Each curve gives
the color excess for an increasing absorption of the screen and for
a given fraction of stars behind the dusty screen. Points along each
curve are separated by $\rm
\Delta A_V = 1$, and the total extinction ranges from
$\rm A_V = 0$ to $\rm A_V=20$. The two boxes show the range of color excess
along the gas bar and outside the bar.
}
\end{figure*}

\begin{figure*}
\centerline{
\psfig{figure=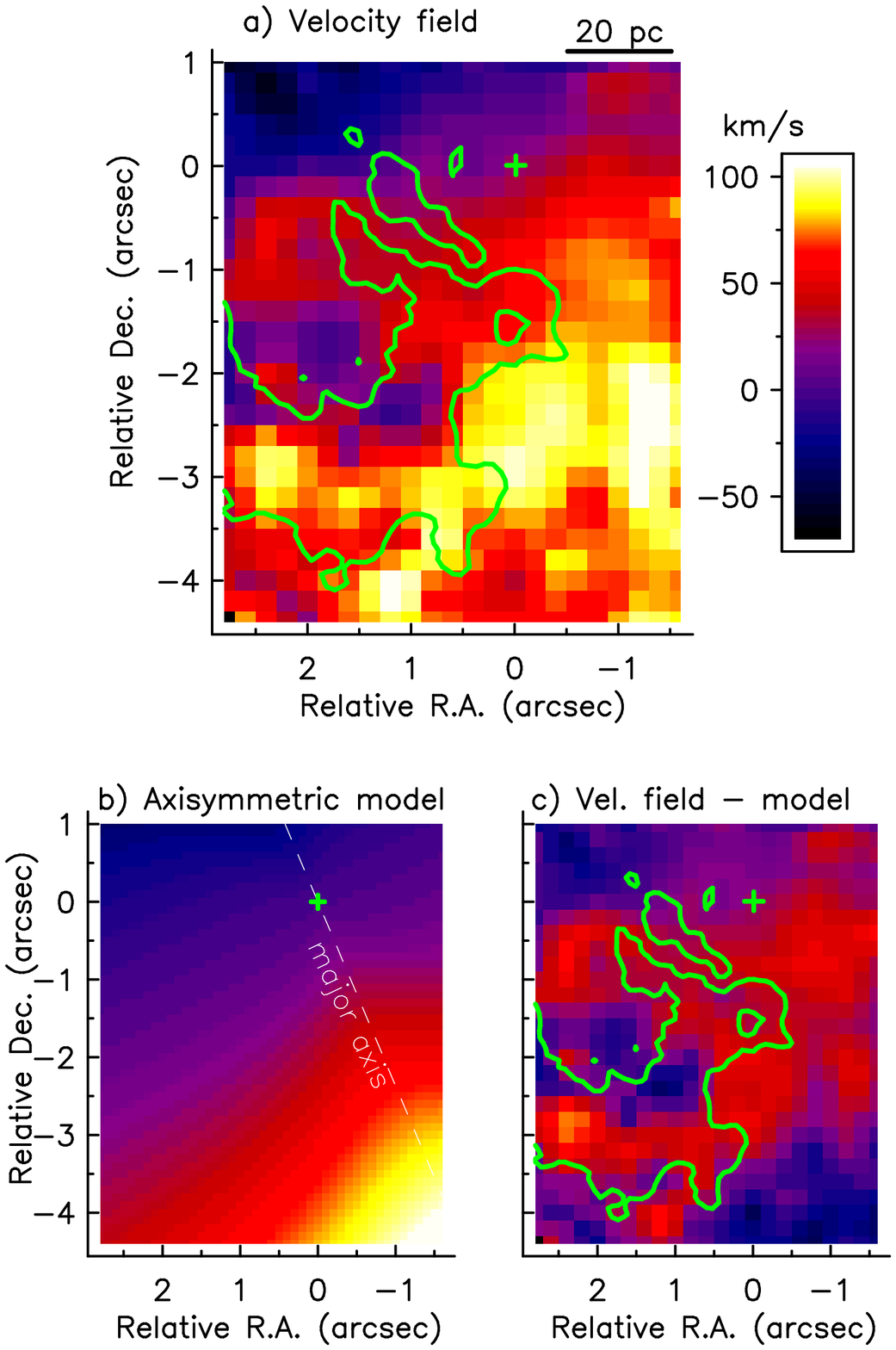,angle=0,width=0.7\linewidth,clip=yes}}
\caption{\label{fig_vfield} 
{\it Color image:} velocity (Doppler) field of the molecular
gas as traced by the H$_2$ S(1)(1--0) emission line at 2.12 $\mu$m,
obtained by means
of the 3D data. {\it Green line:} one of the H--K
contours indicating the location of the dusty bar.
{\it b)} Model of the velocity field expected in the case
of absence of a barred potential (i.e. axisymmetric rotation).
{\it c)} Difference between the observed velocity field and
the axisymmetric model in {\it b}.
}
\end{figure*}

The foreground screen model is the most conservative assumption: given a
color excess it provides the lowest A$_{\rm V}$. In a more realistic case,
dust will be partly mixed with stars. Also, a fraction of stars will also be
present in foreground and their contribution is more important in the
most obscured regions. The latter effect is dominant in the optical, indeed
the heavy dust obscuration affecting the nuclear region
is so high that these regions are essentially invisible
at optical wavelengths. This effect becomes more obvious in Fig.~3 where we
have constructed an R-H color map of the central region of Circinus with the
WFPC2 F606W and the NIC2 F160W images. This map shows a very irregular
structure and no evidence for
the dusty bar detected in the H-K color map. This is a consequence of the
heavy obscuration that
completely absorbs the optical light and that, therefore, makes the R band image
completely dominated by the foreground stars and more affected by small
variations in the extinction. We tried to describe these
effects with a simple but more quantitative model. We make the simplistic
assumption that the distribution of dust and stars can be described by
a homogeneous
dusty screen that hides most of the stars while a small fraction of the
stars is in the foreground. This simple model was used for other heavily
obscured systems (eg. McLeod et al. 1993).
The whole system is then obscured by a Galactic
absorption $\rm A_V (Gal) =1.5$ mag (Freeman et al. 1977).
In this model, the color excess
between two wavebands at $\lambda 1$ and at $\lambda 2$ is given by

\begin{equation}
\rm \matrix{  \rm E_{\rm \lambda 1- \lambda 2}  = 
A_{\lambda 1} (Gal)-A_{\lambda 2} (Gal)+ ~~~~~~~~~~~ \cr
~~~~~~~~~~~~
 - 2.5 log \left[ \rm 1-f_{hid}(1-10^{-A_{\lambda 1}(scr)/2.5})\over
\rm 1-f_{hid}(1-10^{-A_{\lambda 2}(scr)/2.5})\right]   }
\end{equation}

where $\rm f_{hid}$ is the fraction of stars that are hidden by the dust
screen, while A(Gal) and A(scr) are the Galactic and screen absorption
respectively. Fig.~4 shows the expected H--K and R--H color excess as a
function of $\rm A_V(scr)$ and for different fraction of stars behind the
dusty screen.
Two boxes indicate the color excesses measured along and outside the gas bar.
The R--H color excess is the more uncertain quantity, not only bacause of
the observed large dispersion (Fig.~3),
but also because the unabsorbed, intrinsic
R--H color is not well defined. We adopted $\rm (R-H)_0 \simeq 2.3$ from an
average of the colors of Sc spirals in Frogel
(1985) and in Fukugita et al. (1995), but the scatter is very
large. Fig.~4 shows that $\sim$98\% of the stars behind the dusty screen
and a screen absorption of $\rm A_V(scr)\simeq 12$ mag can explain the color
excess observed along the gas bar. The regions next to the bar can
be explained by a fraction of stars behind the screen of $\sim$97\% and
an extiction reduced to $\rm A_V(scr)\simeq 7.5$ mag.

\begin{figure*}
\centerline{
\psfig{figure=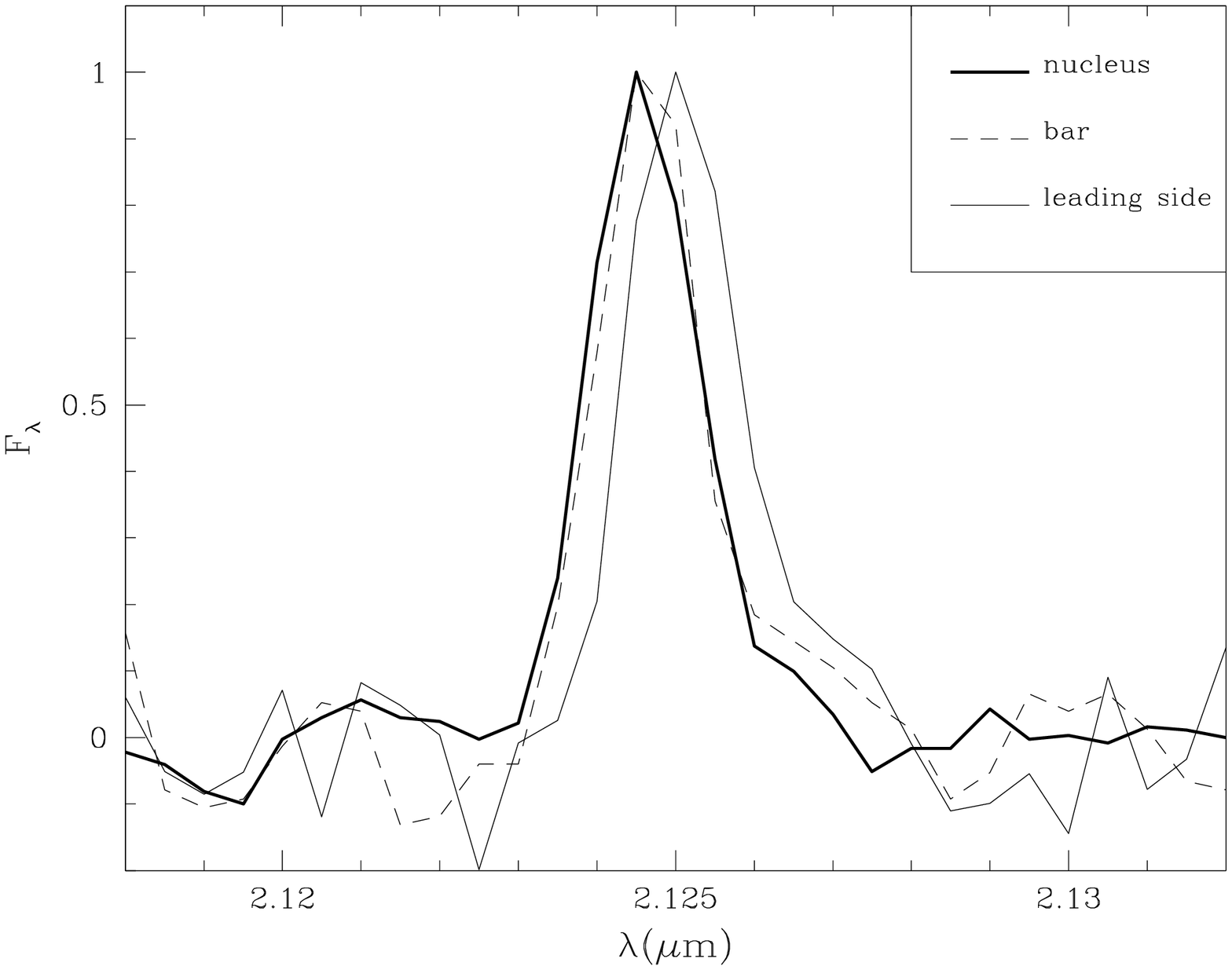,angle=0,width=0.5\linewidth,clip=yes}}
\caption{\label{fig_lineprof} 
Profile of the $\rm H_2(1-0)2.12\mu m$ line on the nucleus,
on the gas bar (1$''$ to the east and 2.2$''$ to the south
of the nucleus) and on the
leading side (0.3$''$ to the west and 2.5$''$ to the south). The extraction
radius is 1$''$.
}
\end{figure*}

If we adopt an intermediate model with $\rm f_{hid} = 97.5$\% for both the
regions along and outside the gas bar, then we expect the following
differences in terms of surface brightness (mag/arcsec$^2$) between the
two zones: $\Delta \mu _R \approx 0.27\pm 0.2$,
$\Delta \mu _H \approx 0.65\pm 0.35$ and $\Delta \mu _K \approx 0.4\pm 0.2$.
Observationally, we obtain: $\Delta \mu _R \approx 0.37\pm 0.22$,
$\Delta \mu _H \approx 0.30\pm 0.10$ and $\Delta \mu _K \approx 0.15\pm 0.04$.
There is a fair agreement between observations and expected values
that, however, is worse at longer wavelength. This probably indicates that
our model is too simplistic and, in particular, that a fraction of star
is probably mixed with the dusty screen and that the dusty screen is not
completely homogeneous as assumed.

The visual extinction along the bar is about 12 magnitudes.
If we assume the Galactic gas-to-dust ratio
($\rm N_H = 2.2\times 10^{21} ~A_V~cm^{-2}$), then the column
density is $\rm N_H \approx 2\times 10^{22} cm^{-2}$. Note, however, that
for active galaxies the standard conversion factor very likely underestimates
the real N$_H$, as discussed in Maiolino et al. (1999b).\\

\subsection{Evidence for streaming motions} \label{3d_bar}

The kinematics of the gas in the presence of a barred potential has been
studied by several authors (Athanassoula 1992, Piner et al. 1995,
Wada \& Habe 1995, Friedli \& Benz 1993). Either gravitational torques due
to the offset of the gas lanes with respect to the bar axis, or shocks
along the gaseous lane, cause the gas on the leading side of the bar
to lose its angular momentum and, as a consequence, to flow toward the
nucleus.  These motions translate into a strong velocity gradient across
the leading side of the bar that should be observed in
the Doppler maps of the emission lines. Such a velocity pattern
has been observed in several large scale stellar bars (eg. Laine et al. 1999,
Regan et al. 1997). We have used our integral field near IR data to trace
the velocity field of the molecular gas with the H$_2$ S(1)(1--0)
emission line at 2.12 $\rm \mu m$. Fig.~5
gives the radial velocity of the molecular gas as traced by
the centroid of the H$_2$ line and relative to the systemic velocity of
438 km/s, derived from the 21 cm line\footnote{The nuclear velocity of H$_2$,
as well as of other low ionization lines and of the stellar features (Fig.6
of Maiolino et al. 1998), shows a small
 redshift of about 20 km/s. This redshift
was also observed in high resolution echelle
spectroscopic observations (Oliva,
private communication). This redshift might arise from the definition
of the velocity for the 21 cm line.
Otherwise, it might reflect small, systemic
differences between the central
region and the outer parts of the galaxy.}.
 The color bar on the right-hand side
of Fig.~5 translates the color coding into the radial
velocity corresponding to the shift of the line. The green curve is one
of the contours of the H--K map indicating the location of the gas bar. 
We excluded from our velocity mapping most of the
regions that were poorly sampled by the dithering (i.e. with low effective
integration time) since the signal to noise on the H$_2$ line is too low
in most of these regions.

The most prominent features of this map are the strong velocity gradient
and  the highly redshifted gas along the leading side of the bar,
indicating gas streaming toward the nucleus in agreement with the model
predictions. This velocity field not only supports the reality of the
gaseous bar, but directly shows inflow motions that probably contribute
to the feeding of the active nucleus.

The gas velocity pattern is similar the what observed in large scale stellar
bars. Here the main difference, besides the much reduced size, is that
the mass that exerts the torques generating the gas inflow is not dominated by
the stars but by the gas.

To show that the redshifted gas
along the leading side is not due to noise or statistical fluctiations,
Fig.~6 shows the
spectrum of the H$_2$ line along the leading side
and compares it with the spectrum of the nucleus and with
the spectrum of the central part of the bar (see figure caption for
details).
The line peak was normalized to one.
 The signal-to-noise on the line is high, and the shift of the line along
the leading side is cleary systemic and not due to noise fluctuations.

The streaming motions become obvious when compared to the rotational
velocity field expected to dominate the kinematics in the absence of a barred
potential. We have modeled the velocity pattern expected if 
the gravitational field is dominated by the stars that are traced by the
K-band light (corrected for extinction), characterized by an average
mass-to-light ratio M/L$_{\rm K} \approx 2.5$ M$_{\odot}$/L$_{\odot}$
(Maiolino et al. 1998). The resulting velocity field is shown in
Fig.~5b.
In the outer regions ($>100$ pc, where the bar should
not play a role), it is in agreement with the observed velocity field (Maiolino
et al. 1998, Storchi-Bergman et al. 1999). Fig.~5c
shows the deviations of the observed velocity field relative to the
rotational field expected in the axisymmetric case: the most obvious
deviation is the strip of redshifted gas along the leading side of the
gaseous bar, in agreement
with model predictions and as observed in large scale stellar bars.

The velocity maps (both Figs. 5 {\it a} and {\it c}) also
show  a strong gradient along the dusty spiral arm that extends from the
southern end
of the bar to the east, also as predicted by models (see eg. Piner
et al. 1995). There is also a region of redshifted gas to the east of the
nucleus; some numerical simulations show the formation of a velocity gradient
in this region (Athanassoula 1992, Piner et al. 1995): indeed this region
is located about at one of the aphelia of the bar, where a significant
population of x$_2$ orbits can reduce the average velocity of the gas
with respect to the circular axisymmetric motions.

The largest velocities of the streaming gas along the bar are expected
to occur close to the center (near the perihelia), as is actually observed
in large scale stellar bars. Such high velocity gas is also observed in
the nuclear region of Circinus. In Fig.~7a the green
contour is a channel map of the H$_2$ emission with velocity
Doppler shift between +200 and +260 km/s, while the color image is
the [SiVI] map that will be discussed in Sect. \ref{cone}.
There is clear evidence for highly redshifted gas at the end of the
gaseous bar, close to the nucleus. In this region the deprojected velocity
is about 300 km/s.

The velocity field of the central region (20--40 pc) is however dominated
by a normal rotation pattern: here the isovelocity lines are parallel
to the minor axis. This nuclear rotation might be associated with the disk-like
feature observed in the H--K color map.

Summarizing, the molecular gas velocity field shows clear non-circular
motions. The general behavior of the kinematics can be ascribed to streaming
motions due to the gaseous bar. Obviously, we cannot rule out more complex
situations, especially concerning irregularities and details of the
velocity field. Higher angular and spectral resolution and a more detailed
modelling of the gas kinematics in this region are required to further
support our findings and/or provide new insights on the details of
the gas dynamics.\\

\section{The innermost region of the ionization cones} \label{cone}

\subsection{The 10 pc scale} \label{10pc_cone}
The gas inflow through the bar not only brings material close to the
nuclear black hole, but probably plays an important role also in other
phenomena of the nuclear activity such as the ionization cones and the
star formation.
Fig.~7a shows the NICMOS map of the [SiVI] coronal emission
line. The [SiVI] emission extends in the ionization cone traced by the
[OIII] optical images (Marconi et al. 1994a), as indicated by the dashed
lines. The presence
of [SiVI] emission also on the eastern side of the nucleus provides evidence
for a counter-cone, whose existence was also inferred from the larger scale
radio maps (Elmouttie et al. 1995, 1998c).
The foreground extinction due to the nuclear dusty
feature (disk?) is very likely responsible for the reduced [SiVI]
brightness of the counter cone. If the nuclear dusty feature produces
an extinction of
A$_{\rm V}\sim 9$ mag (as inferred from Fig.~1b),
this would imply A$_{\rm [SiVI]}\sim 1.1$, fully accounting 
for the brightness difference between the two cones, about a factor of 3.

\begin{figure*}
\centerline{
\psfig{figure=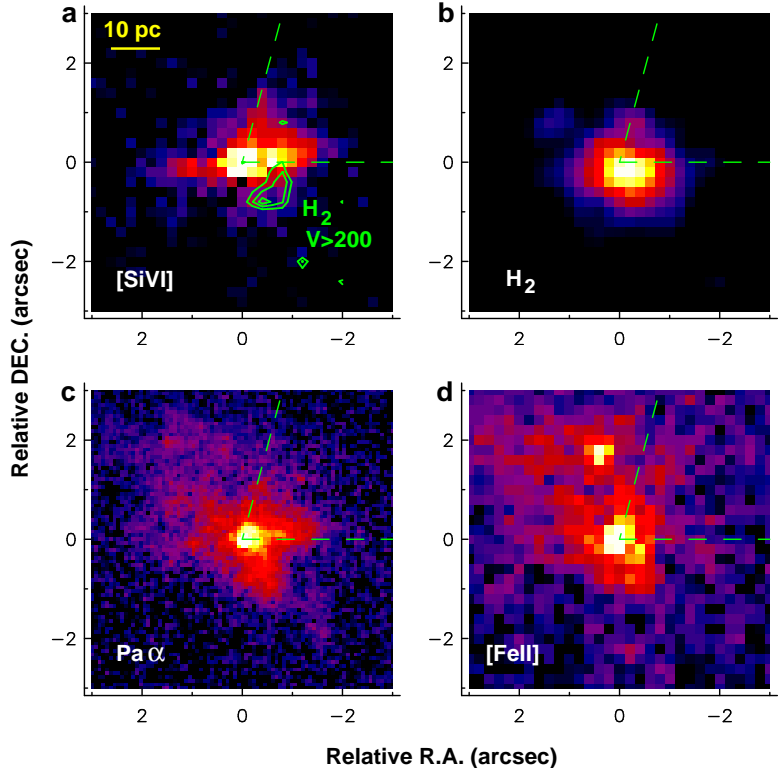,angle=0,width=0.55\linewidth,clip=yes}}
\caption{\label{fig_lines} 
{\it a)} NICMOS map of the [SiVI]1.96 $\mu$m line (color)
along with the 3D channel map of the H$_2$ 2.12 $\mu$m line with the
highest velocity component ($\rm 200 < V < 260$ km/s).
{\it b)} 3D map of the integrated emission of the H$_2$ 2.12$\mu$m
emission line. {\it c)} NICMOS map of the Pa$\alpha$ emission line.
{\it d)} NICMOS map of the [FeII] 1.64$\mu$m emission line. The green
dashed lines indicate the orientation of the light cone observed in [OIII].
}
\end{figure*}

The [SiVI] emission is not azimuthally uniform within the (NW) ionization
cone; the southern edge appears much brighter than the rest. It is likely
that the gas transported by the bar enters the ionization cone in this
region. This
interpretation is supported by the distribution of the molecular gas shown
in Figs. 7 {\it a} and {\it b}, where we report the 3D maps
of the integrated H$_2$S(1)(1--0) emission line and the highest velocity
component of the gas traced by this line. These maps show the presence
of large quantities of gas just south of the ionization cone, that is
flowing at high velocities
towards the southern edge of the ionization cone. Next to the edge of the cone
the H$_2$ emission is highest because here the molecular gas is more
exposed to the nuclear radiation, but also because the gas density is
high and the H$_2$ has not had time to dissociate fully. Indeed, the
R--H color map (Fig.~3) clearly shows a prominent dust lane
extending along the southern edge of the ionization cone, coincident
with the H$_2$ emission.
As soon as these dense clouds enter the ionization cone they are ionized and
emit the [SiVI] line, but also Pa$\alpha$ (Fig.~7c);
this gas is very dense and therefore the ionized gas emission
is very high. Another dust lane in Fig.~3 traces
the northern edge of the ionization cone.

Another possible explanation of the emission line and dust lane morphology
is that these features are simply the effect of the orientation of the
ionization cone with respect the nuclear part of the galactic gaseous
disk, which might be warped on these small scales (Quillen et al. 1999,
Mulchaey et al. 1996, Schreier et al. 1998). The brighter part of the
ionization cone might correspond to the region closer to denser gas of
such nuclear disk. The only difficulty with this model is to obtain a
projected velocity in excess of 200 km/s, as observed in Fig. 7a.

As shown in Figs. 7 {\it a} and {\it b} the molecular
clouds enter the ionization cone very close ($\sim 10$ pc) to the nucleus,
where the radiation pressure is very high. As discussed in Binette (1998)
the radiative acceleration is given by 
$\rm  a_{rad} \approx ({\it n}/1200 cm^{-3})
 (U_0/0.125)(3.1\times10^{-7}+5.8\times 10^{-6}
  \mu _D)~[cm~s^{-2}]$,
where {\it n} is the gas numerical density,
$\rm U_0$ is the ionization parameter
and $\rm \mu _D$ the dust content with respect to the solar value.
The density of the coronal line region was estimated to be about 5000
cm$^{-3}$ based on the infrared [NeV] doublet (Moorwood et al. 1996).
Oliva et al. (1999) determined $\rm U_0$ for a single
cloud in the ionization cone; we scaled this value
to the distance of 10 pc and to the density of 5000 cm$^{-3}$, i.e. $\rm U_0\approx 1$. With these values $\rm a_{rad}\approx 3\times10^{-5}
cm~s^{-2}$ for a dust content of $\rm \mu _D =0.1$. Maiolino et al.
(1998) set an upper limit to the mass of a nuclear black hole in Circinus
of $\rm 4\times 10^6 M_{\odot}$. The gravitational pull at a distance of
10 pc is estimated to be $\rm 4\times 10^{-6} cm~s^{-2}$, i.e. about
one order of magnitude lower than
the radiative acceleration. Therefore, these clouds just driven into the
ionization cone will be pushed outward to form an outflowing component
of the Coronal Line Region and of the NLR. This outward acceleration of
the coronal line clouds, as well as the tendency of the clouds to be
ablated under the strong radiation pressure (Mathews 1986), is probably
responsible for the lower emissivity of the [SiVI] line in the northern
part of the cone.  The denser clouds injected into the cone will survive
ablation and will be accelerated like ``bullets''. Veilleux \& Bland-Hawthorn
(1997) discovered the presence of supersonically ejected clouds
($\rm V_{eject}> 100~km~s^{-1}$) in the Circinus NLR, which they ascribe
to some explosive nuclear event. We claim that such ejected ``bullets''
are the result of the radiative acceleration of the dense clouds that
have been driven into the ionization cone by the gaseous bar or by the
nuclear disk. As suggested by Binette (1998), the radiation pressure--driven
outflow would explain the blueshift of the coronal lines observed in Circinus
(Oliva et al. 1994) and in other Seyfert galaxies (Penston et al. 1984),
as well as the dependence of the blueward asymmetry of the narrow lines
on the ionization potential and on critical density (DeRobertis \& Shaw
1990, Dahari \& De Robertis 1988, Whittle 1992, Heckman  et al. 1981).
A potential problem of the ``outflow model'' would be that new gas must
be continuously supplied to the inner part of the cone
to replace the outflowing gas.  The gas
flow provided by the gaseous bar or by the nuclear disk
could to be the mechanism responsible
for feeding the inner part of the ionization cone (a more quantitative
discussion is given in sect.\ref{fuel_cone}).

On the other hand, a large fraction of
the gas in the outer regions of the NLR (100 pc scale), characterized by
lower ionization emission species, is probably
part of the normal galaxy disk illuminated by the nuclear source. Indeed,
the kinematics of the outer parts of the ionization cone are also
characterized by a rotational component (Elmouttie et al. 1998b,
Veilleux \& Bland-Hawthorn 1997).\\

\subsection{The parsec scale} \label{1pc_cone}
Another interesting feature of the emission line maps in Fig.~7
is that both the [SiVI] and [FeII] lines peak on the nucleus:
although both show extended
emission, a large fraction of the integrated flux ($\sim$ 70\% for [SiVI])
comes from an unresolved source coincident with the 2.2$\mu$m continuum
peak, i.e. the location of the active nucleus.
The shift of the [SiVI] peak observed in the lower resolution data of
Maiolino et al. (1998) was due to a blend of the nuclear component
with the extended component to the west.
Although, a fraction of the coronal lines might be excited by shocks, most
of the coronal line emission must be ascribed to photoionization (Oliva 1997).
However, according to photoionization models, the [SiVI] is expected
to peak at the outer edge of the He$^{+2}$ Str\"{o}mgren sphere, while
the [FeII] should peak right after the H$^+$ Str\"{o}mgren sphere. With
the luminosity of Circinus and a density of 10$^4$ cm$^{-3}$ (i.e. even
higher than what is estimated through the [NeV] infrared lines)
Oliva et al. (1999) derived that the [SiVI] should peak at about 30 pc
from the nucleus. The [FeII] emission should peak even further out.
There are a few possible explanations of this inconsistency:
\begin{enumerate}
\item As suggested by Oliva et al. (1999), the ionized clouds
are dusty. The ionization parameter in these regions is very high (U$\sim 1$)
and, therefore, the dust competes with the gas in absorbing UV photons thus
reducing significantly the size of the Str\"{o}mgren spheres. This
might be the case for the [FeII] emission. Yet the clouds responsible for
the emission of most coronal lines (including [SiVI]) appear characterized
by little dust content\footnote{Oliva et al.
show that very likely most of the Fe
is in the gaseous form in the coronal line region.} and, therefore, dust
probably plays a minor role in the ionization structure of these clouds.
\item
Another possibility is that the gas that emits most of the coronal lines is
significantly denser, shrinking the size of the He$^{+2}$ sphere.
More specifically $\rm R(He^{+2})\propto N_e^{-2/3}$.
Reducing the radius of the [SiVI] emitting shell from 30 pc
to less than 10 pc (which would make the emission unresolved) would
require an electron density larger by a factor of 5, i.e. $\rm N_e =
5\times 10^4 cm^{-3}$. However, this density is one order of magnitude
higher than what measured from the [NeV] doublet (Moorwood et al. 1996).
Also, the thickness of the [SiVI] emitting shell
(Marconi et al. 1994b), relative to $\rm R(He^{+2})$, is inversely
proportional to N$_e$ and, therefore, the [SiVI] line would be significantly
fainter than other coronal lines that form inside the He$^{+2}$ sphere,
i.e. higher ionization species. This line of reasoning applies
as long as the ionization structure has a quasi-spherical geometry
(Oliva 1997). If a nuclear population of very dense clouds exists, more
specifically with
$\rm N_e > 10^5 cm^{-2}$ (but still lower
than the critical density of $\sim 10^8 cm^{-2}$), then the geometry
of the ionization structure is quasi-plane parallel ($\rm R > 10~\Delta
R$) and the region of [SiVI] emission extends over the whole He$^{+2}$
Str\"{o}mgren region. This would solve the problem of the ratio between
[SiVI] and the higher ionization coronal lines, but the lower density
measured with the [NeV] lines remains an issue.
\item
Increasing the filling factor would also reduce the size of the emission
region ($\rm R(He^{+2})\propto f^{-1/3}$), but this would also affect
the coronal line ratios similarly to the change in density discussed above.
\item
A final, interesting possibility is that the nuclear UV emission is
intrinsically non--isotropic (Oliva et al. 1999), as is expected to be
the case if the emission arises from an optically thick accretion disk.
In the latter case the azimuthal clouds are illuminated by a flux that
is significantly lower
than the face-on flux (L$_{\rm AGN} \propto (\cos \theta )^{-1}$, Laor
\& Netzer 1989). The He$^{+2}$ Str\"{o}mgren radius of the azimuthally
distributed gas
would be much smaller than that estimated above and could very well appear
point like at our resolution.
\end{enumerate}

\section{Nuclear star formation} \label{sf}

The Pa$\alpha$ and the [FeII] emission appear much more diffuse and
extended\footnote{The [FeII] morphology observed in our NICMOS image
is consistent with the lower resolution map obtained by Davies et al.
(1999).} than the [SiVI]. Both emission lines extend well outside the
ionization cone and preferentially along the major axis of the galaxy,
although the Pa$\alpha$ also has a component in the ionization cone similar
to that observed in the [SiVI] map. Very likely the diffuse/extended
emission traces moderate
star forming activity occurring within the central 50 pc as a result of the
large amount of molecular gas driven into this region by the gaseous bar.
The Pa$\alpha$ emission traces the UV radiation from young OB stars. Maiolino
et al. (1998) used the circumnuclear Br$\gamma$ emission and the mass-to-light
ratio to model the star formation history, and infer that the nuclear stellar
population has an age of about 10$^8$ yr. 

Fig.~7c shows that the star formation history of the nuclear
region is probably more complex. The Pa$\alpha$ map shows the presence
of knots and non-axisymmetric emission: a chain of three HII region is
present at
about 20 pc to the north-east of the nucleus, while a burst of star formation
is present to the south-west, right at the nuclear end of the
leading side of the gaseous bar. These are the locations of the aphelia
of the bar, where orbit crowding is indeed expected
to enhance the star formation (Wada and Habe 1992). There is no evidence
for star forming activity along the bar. Although large concentrations
of gas are expected along the leading side of the bar, star formation might
be inhibited by the strong shear forces (Elmegreen 1979, Kenney \& Lord 1991).

The diffuse [FeII] emission probably traces gas shocked by supernova
explosions.  A bright [FeII] knot, about 30 pc
to the north of the nucleus very likely traces a single young supernova
remnant. The [FeII] knot is unresolved (i.e. R$<$ 4 pc), thus implying that
the SNR must be younger than about 100 yr. There is also a prominent [FeII]
extension to the south of the nucleus, slightly offset with respect to the
Pa$\alpha$ emission; shocks due to the gas streaming from the gaseous bar
might also play a role in this region.

\section{Discussion} \label{disc}

\subsection{Plausibility of the gas bar and open issues} \label{issues_bar}

Although the presence of a nuclear gas bar appears convincing both from the
morphological and the kinematical points of view, we can check whether
the physical conditions of the gas meet the requirements necessary
to develop the gravitational instabilities that should form a bar.
According to models, a gaseous bar should form when the mass of the gaseous
nuclear disk is larger than 10 (Wada \& Habe 1992) to 20\% (Shlosman
et al. 1989), of the dynamical mass. CO(2--1) millimetric data (Aalto et al.
1995, Johansson et al. 1991) indicate that the gas mass enclosed within
the central 400 pc is about $\rm 6 \times 10^8 M_{\odot}$, if the
Galactic conversion factor is assumed (Kenney \& Young 1989). Based on
the kinematical information provided by Maiolino et al. (1998) we estimate
that the dynamical mass inside the same region is about $\rm 1.2 \times
10^9 M_{\odot}$, i.e. only a factor of two higher than the mass in
molecular gas. Therefore, the physical conditions of the gas in the
nuclear region are indeed appropriate for the formation of a gaseous bar
(although this argument is weakened if the conversion from CO to gas mass
is less than the standard Galactic value, as sometimes appears to be the
case in extreme starbursts).

In the H--K color map (Fig.~1b)
only one side of the gaseous bar is observed;
the north--west counterpart is not seen. One possible explanation is that
the other side of the gaseous bar is present but it is not observed in the
extinction map because the contrast of absorption features is much
lower on the far side of the disk (more foreground stars and less stars
behind). Otherwise the north-west side of the bar might
be intrinsically weaker or even absent. Since gaseous bars are formed via
gravitational instabilities, they are expected to be highly
inhomogeneous and possibly asymmetric, especially on such small scales.

Another important issue to address is whether these gaseous bars are
rare cases or they could be a fuelling mechanism common to most active
galaxies as claimed by Shlosman et al. (1989). If the latter is the case
then the question is why such nuclear gaseous bars were detected only
in a few active nuclei (Mulchaey \& Regan 1999, Henkel et al. 1991,
Ishizuki et al. 1990), or none if we only consider nuclear bars as small as
the one in Circinus. These gaseous bars can be identified from
molecular gas emission (eg. CO millimetric maps) and/or through dust absorption traced by the color maps, as we have done in this work.
Circinus is closest Seyfert 2 known and the projected angular size of
the gaseous bar is 2 arcsec. Most of the other Seyfert galaxies are
further away and, therefore, a nuclear gaseous bar with similar size
would have a
much smaller angular size, which would make it hard to detect.
The average distance of the Seyfert galaxies in the sample of Maiolino
\& Rieke (1995) is 32 Mpc and the projected size of the nuclear gaseous
bar would be a few tenths of arcsecond, unresolved even at the NICMOS
resolution (at 2 $\mu$m) and at the resolution of the current millimetric
interferometers. Also, we
emphasize that the color map technique to search for gaseous bars works as
long as infrared filters are used. In these heavily obscured nuclei the
optical images sample only the outer (foreground)
 parts of the nuclear region and therefore
do not provide the required information even if compared with an infrared
image to obtain a color map. Indeed, as shown in Fig.~3 and as discussed in
sect.3.1, the R--H color map of Circinus does not show
evidence for the dusty bar detected in the H--K color map.
This result may partly explain the shortage of gaseous bars in Regan
\& Muchaey (1999), a work which included images taken through the above
filters of a sample of Seyferts. Summarizing, whether nuclear gas bars
are a common feature in Seyfert galaxies or not remains an open question,
since such
structures might have been missed in previous studies both because of
limited angular resolution and/or inappropriate techniques.

\subsection{Fuelling the active nucleus} \label{fuel_agn}

As discussed in Athanassoula (1992) and Regan et al. (1997), probably only a
small fraction of the streaming gas does actually result in an inflow
(10\%--30\%). We can estimate the streaming flux of the gas along the
leading edge of the nuclear bar in Circinus and compare it with the
inflow requirements to feed the active nucleus. The gas streaming flow
along the leading edge of the bar (an upper limit to the real inflow rate)
is given by:

\begin{equation}
\rm \dot{M}_{stream} = N_H~w~V_{stream}~m_p
\end{equation} \label{m_stream}

where w is the width of the streaming gas lane ($\sim$ 20 pc), m$_p$ is the
proton mass, $\rm V_{stream}$ is the streaming
gas velocity along the leading edge of the gaseous bar and $\rm N_H$ the
gas column density in the same region ($\sim 2\times 10^{22} cm^{-2}$,
sect.\ref{nic_bar}).
To determine $\rm V_{stream}$
we have to correct the observed redshifted velocity for various
projection effects and for the contribution due to the pattern circular
motion, as illustrated in Fig.~2. More specifically:
\begin{equation}
\rm V_{stream} = {V_D(obs)/\cos
{\it i} - V_{pattern} \sin \phi \over \cos \alpha}
\end{equation} \label{v_stream}
 where V$_{\rm D}$ is the Doppler
velocity shift observed along the leading edge of the bar,
$\phi$ is the deprojected angle between the galaxy minor axis\footnote{More
specifically the direction orthogonal to the line of nodes and on
the plane of the galaxy.} and
the line connecting the nucleus to the specific point along the leading edge
of the bar where the redshifted gas is observed (see Fig.~2),
V$_{\rm pattern}$ is the velocity component due to pattern speed at the
same location, while {\it i} and $\alpha$ were defined in Sect.\ref{nic_bar}.
The outer edge of the gaseous bar is located at about
160 pc from the nucleus, probably the location of the corotation
radius. At this distance the rotational velocity
is 150 km/s (deprojected for inclination,
Maiolino et al. 1998), thus giving a pattern speed of
$\rm \Omega_{pattern} = 0.92$ km s$^{-1}$ pc$^{-1}$. If we consider the
streaming gas located at about the mid-point of the gaseous bar, then
the deprojected distance from the nucleus is about 90 pc, hence V$_{\rm
pattern} =
83$ km/s, and $\phi = 25^{\circ}$. The contribution from the circular pattern
velocity ($\rm V_{pattern} \sin \phi$) is very low, as expected since the
bar is oriented almost along the line of sight. With these values
we estimate $\rm V_{stream} = 83$ km/s, hence from Eq.
\ref{m_stream} we derive a streaming flow rate of $\rm \dot{M}_{stream}
\approx 0.25~M_{\odot}yr^{-1}$.

The luminosity of the active nucleus is about $2\times
10^{10}L_{\odot}$, or lower if the nuclear emission is intrinsically
anisotropic (Oliva et al. 1999), thus implying an accretion rate of
\begin{equation}
\rm \dot{M}_{AGN}
= L_{AGN}/c^2\epsilon \approx 10^{-2} M_{\odot}yr^{-1}
\end{equation}
(where we assumed a radiation efficiency $\epsilon = 0.1$).
The streaming flow estimated above is 25 times higher than the AGN accretion
rate. Therefore even if only 10\% of the streaming gas actually flows
into the nucleus, this will provide 2.5 times as much fuel as required
to power the activity.

\subsection{Fuelling the nuclear star formation} \label{fuel_sf}

As discussed above, Pa$\alpha$ emission traces regions of recent star
formation, while the smoother near-IR continuum emission
traces the older stellar population evolved from past bursts.
We can estimate the current star formation rate from the
Pa$\alpha$ emission due to stellar activity integrated in the nuclear region.
We subtracted the contribution to the Pa$\alpha$ emission due to the
active nucleus by normalizing the [SiVI] map to the Pa$\alpha$ on the
nucleus; this is actually a conservative subtraction of the AGN Pa$\alpha$
flux, since a fraction of the unresolved nuclear Pa$\alpha$  emission, as
well as part of the Pa$\alpha$ emission within the cone, might be due
to star formation as well. The flux of the AGN-subtracted Pa$\alpha$ integrated
within a radius of 30 pc ($=1''.5$) is $\rm 1.2\times 10^{-13}$ erg s$^{-1}$
cm$^{-2}$. If we assume that the extinction derived from the H--K color
map in Fig.~1 also applies to the nebular emission, then
on average A$_{\rm V}\sim 9$ mag, hence A$_{\rm Pa\alpha}\sim 1.3$ mag.
Therefore the extinction corrected Pa$\alpha$ flux is $4\times 10^{-13}$
erg s$^{-1}$ cm$^{-2}$. At the distance of 4 Mpc, this flux translates
into a ionizing photon flux $\rm Q_i = 1.4\times 10^{51} ~s^{-1}$. We
employed the stellar population synthesis code described in Sternberg
\& Kovo (1999) to derive the star formation rate required to produce
this ionizing photon flux.
By adopting a Salpeter initial mass function and an upper mass cut-off
of 60 M$_{\odot}$ we obtain a star formation rate of $\rm
\sim 10^{-2} ~M_{\odot} yr^{-1}$. This star formation rate is relatively
low and similar to the accretion rate of the active nucleus. The amount
of gas inflow provided by the gaseous bar (sect.\ref{fuel_agn})
can feed both the active nucleus and the star forming
activity in the nuclear region.

\subsection{Fuelling the ionization cone} \label{fuel_cone}

As discussed in sect.\ref{10pc_cone}
part of the gas observed in the ionization cone is
in circular motion in the galactic plane, while a fraction of the gas
is outflowing, probably as a consequence of the nuclear radiation pressure.
We speculated that the inflow due to the gas bar might
supply to the nucleus the material that is being ejected through the
ionization cone. Here we address this possibility more quantitatively.
Veilleux \& Bland-Hawtorn (1998) estimate the kinetic energy in the outflowing
gas and in the ejected filaments to be about $\rm K\approx
10^{52}(N_e/100cm^{-2}) ~erg$. By approximating with a spherical outflow,
the kinetic energy can be converted into a mass outflow rate according
to the formula $\rm \dot{M}_{outfl} \approx 2K/(V_{outfl}R) $, where
$\rm V_{outfl}$ is the (deprojected) outflow
velocity and R is the average distance of the outflowing gas detected
in the images. By assuming $\rm V_{outfl}\approx 200$ km/s (Veilleux \&
Bland-Hawtorn 1998) at an average distance of about 100 pc, and by
assuming an average density $\rm N_e \approx 500 cm^{-3}$ (Oliva et al. 1999),
we obtain $\rm \dot{M}_{outfl}\approx 10^{-2} M_{\odot}yr^{-1}$. Although
this estimated is very rough, it shows that the outflowing component of the
ionization cone requires some degree of fuelling, and that the inflow
due to the gas bar can account for this fuelling.

\section{Conclusions} \label{concl}

We present NICMOS/HST images and integral field near-IR spectroscopy of the
nuclear region of the closest Seyfert 2 galaxy Circinus.
The H--K color map shows a nuclear dust lane that extends radially.
We propose that this dust lane traces a nuclear gas bar that might be
responsible for feeding the active nucleus. The velocity field traced
by the H$_2$ line at 2.12 $\rm \mu m$ shows enhanced radial motions along
the leading edge of the gas bar, in agreement with model predictions.
The latter finding not only supports the
identification of the gaseous bar, but directly shows gas inflow motions
that probably fuel the AGN. The estimated streaming flow along the leading
edge of the bar is about $\rm 0.25~M_{\odot}yr^{-1}$. Probably no more than
10--30\% of this flow does actually result in an inflow rate, that is
however enough to fuel the active nucleus (whose accretion rate is about
$\rm 10^{-2}~M_{\odot}yr^{-1}$). We cannot rule out more complex situations
to explain details and irregularities of the velocity field;
higher resolution spectroscopic data are required to further
support our findings and/or provide new insights on the details of the
gas dynamics.

Nuclear gaseous bars similar to the one
in Circinus might be a fuelling mechanism common to many
active nuclei hosted in gas rich galaxies.
Possibly, limited angular resolution and/or inappropriate
techniques might be responsible for the shortage of nuclear gas bar detections
in other active nuclei. 

We also used the NICMOS data to study the innermost part of the Narrow Line
Region and the Coronal Line Region. The coronal line [SiVI] 1.97 $\mu$m
reveals the presence of a counter-cone that was not observed at optical
wavelengths, very likely because of obscuration. The [SiVI] and Pa$\alpha$
emission is not distributed uniformly within the ionization cone but
extends preferentially along the southern edge of the cone. This enhanced
emission seems associated with dense gas just outside the cone, traced
by the H$_2$ (molecular gas emission) and R--H (dust absorption) maps.
This gas is flowing towards the cone at velocities in excess of 200 km/s.
We speculate that the gas streaming from the gas bar enters the ionization
cone, contributing significantly to the Coronal Line Region and to the
innermost part of the Narrow Line Region.
This dense gas is probably accelerated by the radiation pressure producing the
outflowing gas and the ejected clouds observed in the Narrow Line Region.

Besides the extended component, both the [SiVI] 1.97 $\mu$m and the [FeII]
1.64 $\mu$m emission are characterized by a nuclear intense unresolved
component, while photoionization models would expect these lines to
peak at a few 10 pc from the nucleus. This finding poses important
constraints on the photoionization models.

Both Pa$\alpha$ and [FeII] emission extend over the central 60 pc and
also outside the ionization cone. Most probably this extended emission
traces recent star formation in the nuclear region. The Pa$\alpha$
extended emission is irregular and ``knotty'', different than the
smoother near-IR continuum. This Pa$\alpha$ emission
probably traces recently formed HII regions, therefore
pointing to a complex nuclear star formation history. We estimate an average
star formation rate in the nuclear 60 pc of $\rm \sim
10^{-2}~M_{\odot}yr^{-1}$, which could be sustained by the inflow rate
doe to the gaseous bar.

\acknowledgments

We thank E. Oliva, A. Marconi and the referee, M.W. Regan,
for useful comments.
During the course of this work A.A-H. and A.Q. were
supported by the National Aeronautics and Space Administration (NASA)
on grants NAG 5-3042 and GO-07869.01-96A respectively,
through the University of Arizona.
R.M. acknowledges the partial financial support
from the Italian Space Agency (ASI)
through the grant ARS--98--116/22 and from the
Italian Ministry for University
and Research (MURST) through the grant Cofin98-02-32.

\end{document}